\documentclass[reprint,aps,prl,twocolumn,superscriptaddress,showpacs,floatfix,10pt]{revtex4-1}
\usepackage{graphicx}
\usepackage{amsmath}
\usepackage{amsfonts}
\usepackage{amssymb}
\usepackage{mathrsfs}
\usepackage{longtable}
\usepackage[rightcaption]{sidecap}
\sidecaptionvpos{figure}{t}

\begin{document}

\title{Growth and instability of a phospholipid vesicle in a bath of fatty acids}

\date{\today}

\author{Julien Dervaux}
\affiliation{Center for Studies in Physics and Biology, Rockefeller University, 1230 York Avenue, New York, NY 10021, USA}
\author{Vincent Noireaux}
\affiliation{School of Physics and Astronomy, University of Minnesota, Minneapolis, MN 55401, USA}
\author{Albert Libchaber}
\affiliation{Center for Studies in Physics and Biology, Rockefeller University, 1230 York Avenue, New York, NY 10021, USA}

\begin{abstract}
Using a microfluidic trap, we study the behavior of individual phospholipid vesicles in contact with fatty acids.  We show that spontaneous fatty acids insertion inside the bilayer is controlled by the vesicle size, osmotic pressure difference across the membrane and fatty acids concentration in the external bath. Depending on these parameters, vesicles can grow spherically or become unstable and fragment into several daughter vesicles. We establish the phase diagram for vesicle growth and we derive a simple thermodynamic model that reproduces the time evolution of the vesicle volume. Finally, we show that stable growth can be achieved on an artificial cell expressing a simple set of bacterial cytoskeletal proteins, paving the way toward artificial cell reproduction. 
\end{abstract}



\maketitle


The RNA world hypothesis posits that the current protein-based enzymatic system at work in modern lifeforms may have originated from an RNA-based system where catalytic activities were performed by ribonucleic acids \cite{rna,orgel04}. This hypothesis raises fundamental questions such as how these primitive building blocks of life have reached the high concentrations that are necessary for the emergence of life. In this respect, it is thought that RNA compartmentalization  within self-assembled vesicles could have allowed the formation of RNA-rich phases while avoiding the dilution of information that arises in open systems \cite{szostak01,chen10}. Because it is unlikely that the complex machinery controlling modern cell division was at work in protocells \cite{szostak01}, it is reasonable to assume that environmentally-controlled physico-chemical processes may have come into play to trigger protocell growth and division. Experimental studies have indeed shown that fatty acids (FA) vesicles can self-assemble \cite{gebicki73,hargreaves78}, grow \cite{hanczyc03} and divide \cite{walde94,zhu09} under plausible prebiotic conditions. Besides the relevance of these results for the origin of life, using the spontaneous ability of vesicles to replicate could also provide a valuable route towards the production of self-replicating artificial cells. However, this objective requires a precise control over the growth mechanism and so far most quantitative experiments on vesicle reproduction have been performed in batch mode at the cell population level using indirect methods to detect average vesicle size and number \cite{walde94,hanczyc03,chen04,chen04b,markvoort10}. While some studies have also investigated FA/vesicles interactions using light microscopy, they have focused on the large spectra of instabilities \cite{peterlin09,zhu09,hentrich14} and the accurate control of spontaneous vesicle growth remains to be achieved. With these considerations in mind, we study in this paper the interactions of individual phospholipid vesicles with a FA bath.



%
%
%
%
%



\begin{figure}[h!]
	\centering
		\includegraphics[width=0.93\columnwidth]{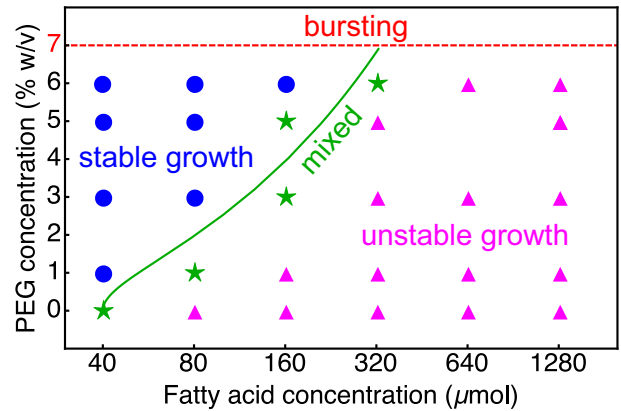}
		\caption{(color online). Phase diagram describing the different regimes of interaction between a PEG-encapsulating GUV and a bath of FA at various concentration. The solid green line is the curve $(c\log{\frac{c}{c_0}})/\phi^2 = 14 \mu$M with $c_0 = 40 \mu$M (see model). All vesicles with internal PEG concentration of $7\%$ exploded following FA addition.}
	\label{fig:phasediagram}
\end{figure}

Giant unilamellar vesicles (GUV) of phosphatydilcholine, encapsulating an internal aqueous phase I (PEG-4500 with concentration from $0$ to $6\%$ w/v and 50 $\mu$M of the fluorescent dye AlexaFluor 488) and immersed in an external aqueous phase II ($6.6$ mM Tris, pH 8.5), were prepared by the inverted emulsion technique \cite{pautot03,noireaux04} (see Supplementary materials).  \begin{figure}[t!]
	\centering
		\includegraphics[width=0.93\columnwidth]{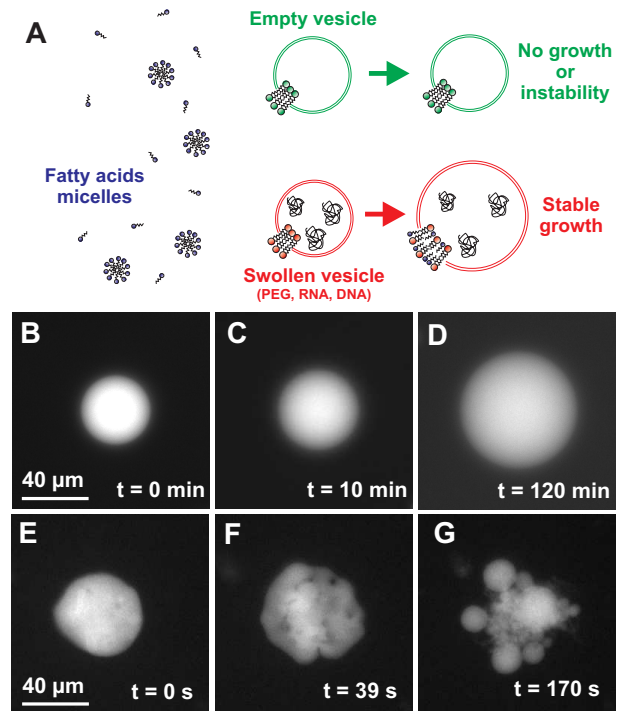}
		\caption{(color online). A: Schematic representation of the experiment. B-D: Stable growth of a vesicle encapsulating $6\%$ PEG immersed in a bath of FA at $160\mu$M. E-F: Unstable growth and formation of daughter vesicles. The inner PEG concentration is $0\%$ and the outer FA concentration is $320\mu$M.}
	\label{fig:growth-illustration}
\end{figure}  To induce vesicle growth, the GUV solution was diluted in freshly prepared oleic acid (Nu-Check) micelles dispersed in phase II (FA concentration $c$ between 40 $\mu$Mol to 1280 $\mu$Mol). A $\sim0.5 \mu$L droplet of the GUV-FA mixture was immediately placed between two coverslips separated by a 250 $\mu$m thick spacer. The shape of large vesicles (diameter $\geq10\mu$m) was monitored over time using an inverted fluorescence microscope (Olympus IX70) equipped with a CCD camera (QImaging Retiga 1300). Representative pictures of the experiments are presented in Fig.\ref{fig:growth-illustration} B-G. Depending on the PEG and FA concentrations, we observed several regimes: i) slow stable growth retaining the spherical symmetry (Fig. \ref{fig:growth-illustration} B-D and movie S1), ii) fast unstable growth with symmetry breaking (Fig.\ref{fig:growth-illustration} E-G and movie S2), iii) coexistence of vesicles undergoing stable or unstable growth and finally iv) vesicles bursting following FA addition. The phase diagram in Fig.\ref{fig:phasediagram} shows the repartition of the different regimes in the [PEG]-[FA] phase space. At high FA concentrations, in the unstable growth regime, fluctuations of the vesicle membrane are visible almost immediately (within $\sim 30$ seconds) following the addition of FA. The amplitude of these fluctuations increases for a few minutes until being comparable to the vesicle diameter at which point the GUV fragments into several stable small daughter vesicles. For the largest FA concentration (1280$\mu$M) and internal PEG $\leq 1\%$, we also observed the formation of long tubular structures which further underwent a pearling instability, as previously reported \cite{hentrich14}. The formation of tubular structures may be due to the generation of a spontaneous curvature in the membrane due to the relatively slow flip-flop rate of the FA molecules. In regime iii) we also observed transient instabilities where the vesicles apparently fragmented into several smaller vesicles before reincorporating them (movie S3). As mentioned in the supplementary materials, vesicles larger than 5$\mu$m and containing $\geq 7\%$ internal PEG were hard to produce and were typically small due to the large osmotic pressure difference across the bilayer membrane. For such vesicles, when FA were added to the external solution (in regime iv), vesicle bursting occurred. This bursting can be ascribed to the decrease in the membrane tensile strength as a result of the incorporation of FA. 

\begin{figure*}[t!]
	\centering
		\includegraphics[width=0.93\textwidth]{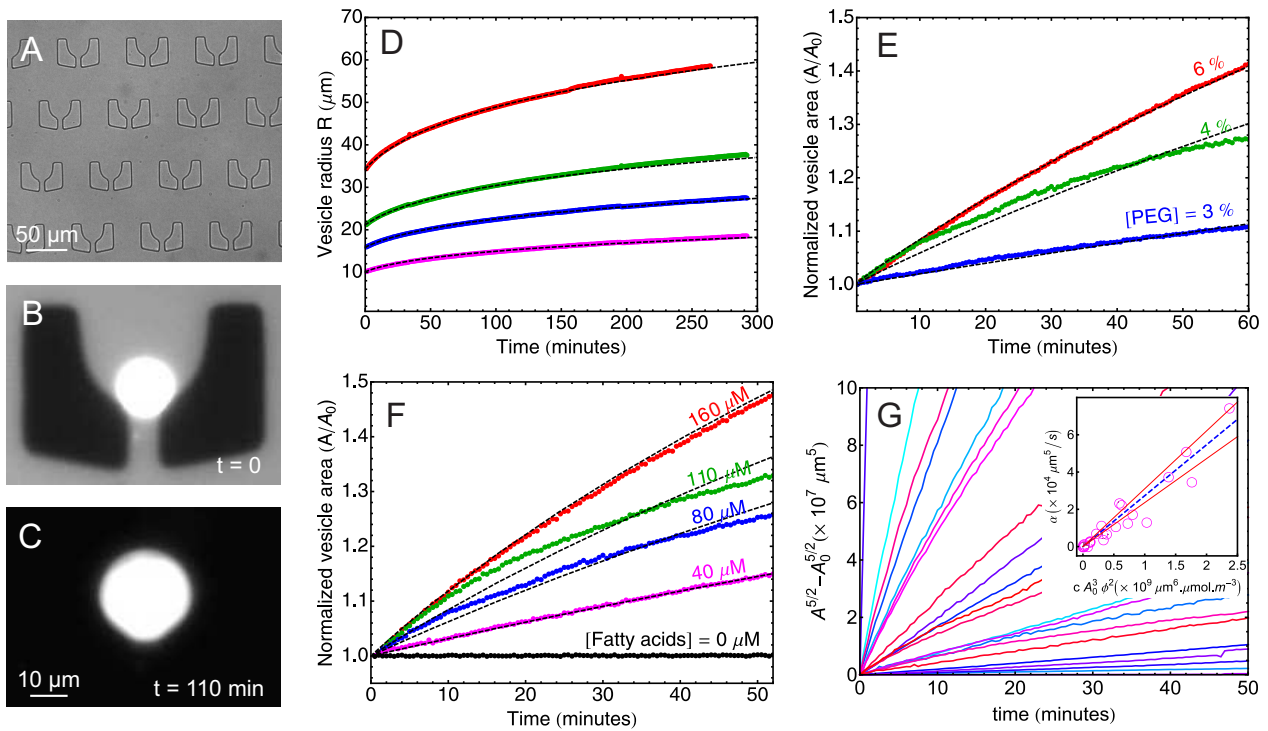}
		\caption{(color online). A: Microfluidic device used to trap phospholipid vesicles. Once a vesicle is trapped (B) the external medium is flushed and FA are flowed around the vesicles which grow (C). D: Time evolution of vesicle sizes for different initial radii. The internal PEG concentration is $6\%$ and the outer FA concentration is 160$\mu$M. E-F: Time evolution of normalized vesicle areas for various internal PEG concentration (E) (FA = 80$\mu$M) and for various outer FA concentrations (F) (inner PEG=$4\%$). Initial radii are $21\pm1\mu$m. G: The rescaled areas $A^{5/2}-A_0^{5/2}$ show a clear linear dependence. The slopes of these curves are proportional to $c A_0^3 \phi^2$ (inset). }
	\label{fig:growth-curves}
\end{figure*}

We then focused on the stable growth regime and, in order to accurately characterize the growth of the vesicles for various physical parameters in a stationary environment, we used a microfluidic trap \cite{huebner09} to immobilize the vesicles under the microscope, as illustrated in Fig.\ref{fig:growth-curves} A-C. Once some vesicles were trapped, the external medium was first flushed with the aqueous phase II to remove traces of unencapsulated materials. Next, FA dispersed in phase II were injected at a constant flow rate inside the microfluidic device using a syringe pump. The flow rate was adjusted to minimize vesicle deformations.  We first investigated the effect of vesicle size on growth. The internal PEG concentration was fixed at $6\%$ and the fatty acids concentration at $160 \mu$Mol. We monitored for 300 minutes the time evolution of the radius of the vesicles (with initial radii ranging from 10 to 35 $\mu$m) and plotted the results in Fig.\ref{fig:growth-curves} D. The radii were found to increase monotonically over the course of the experiments, with up to a five-fold increase of the vesicle volume. With the continuous feeding of FA, no saturation of the growth process was observed. It can also be seen from the slopes of these curves that larger vesicles grow at faster rates than smaller vesicles.  Additionally, we monitored the total fluorescence signal (integrated over the vesicle) and only a decrease of $\leq 5\%$ was measured over the course of the experiment, thus showing that dye leakage from the vesicle was minimal. Given that PEG-4500 has a larger molecular weight  than the fluorescent dye ($721$ g/mol), this indicates that the number of PEG molecules inside the vesicle is well conserved during the growth process. Occasionally, we observed small bursts of fluorescent dye escaping from the vesicle. Because these bursts led to a sharp decrease in vesicle volume, data points after the bursts were discarded. Next, we also investigated the dependance of the growth process on the internal PEG and external FA concentrations. To isolate these effects from the size-dependent growth rate, vesicles with similar diameter (within $5\%$) were selected and their growth was monitored over time. The time evolution of their area $A$ (scaled by their respective initial area $A_0$) is plotted in Fig.\ref{fig:growth-curves} E-F for various values of the control parameters. Increasing the PEG or FA concentrations increased the vesicle growth rate. Growth could also be induced by replacing the internal PEG by single (90 thymine sequence) or double-stranded DNA ($\lambda$-phage genomic DNA).








To analyse these data, we now write a simple thermodynamical model describing the osmotically-driven growth of a phospholipid vesicle of radius $R$ in contact with a reservoir of fatty acids at molar concentration $c$. In all generality, fatty acids (FA) and solvent particles migrates from the external bath to the membrane bilayer or to the vesicle interior if this decreases the Gibbs free energy of the system. If the migration of solvent particles is fast compared to the migration of the FA, the vesicle remain swollen and spherical. On the other hand, if FA incorporation in the membrane bilayer exceeds the permeation of the membrane by solvent particles, the area A of the vesicle may grow faster than its volume V and we can expect the vesicle to become floppy and to lose its sphericity. In order to first describe the stable growth regime, we shall make the assumption that the migration of solvent molecules is fast enough for the vesicle to remain spherical. Under this assumption, the total number N of FA molecules in the membrane is governed by the mass conservation equation: $\partial N/\partial t = A ( J_{ext} - J_{int})$ where $J_{ext}$ and  $J_{int}$ are respectively the flux of FA molecules from the external bath to the membrane and from the membrane toward the vesicle interior. However, because we expect that the FA concentration at equilibrium will be roughly the same inside and outside the vesicle, the total number of FA molecules inside the vesicle (around $10^5-10^6$ for a $20\mu$m diameter vesicle in a 100$\mu$M bath) will be negligible compared  to the total number of FA molecules in the membrane (around $10^9$ for a $10\mu$m diameter vesicle doubling its area and assuming that the area per molecule is $a=30$ {\AA}$^2$ ) and we shall neglect the flux $J_{int}$ in the following. According to Fick's law, the flux $J_{ext}$ is related to the change in chemical potential by $J_{ext} = \mathcal{N} c D/ \ell k_B T (\mu_{ext} - \mu_{ves}) $ associated with the insertion of a FA molecule inside the vesicle membrane. $\mathcal{N}$, $D$, $k_B$ and $T$ are respectively the Avogadro constant, FA diffusion coefficient, Boltzmann constant and temperature.  $\ell$ is a lengthscale that we will discuss later. In the external bath the chemical potential $\mu_{ext} $ of the FA is simply that of an ideal solution. The chemical potential $\mu_{ves}$ is related to the change in Gibbs free energy $G_{ves}$ of the vesicle by $\mu_{ves} = \partial G_{ves} /\partial N$ associated with a variation of the number N of FA in the vesicle membrane. In the limit of fast water transport, $G_{ves}$ contains two contributions: i) a membrane term describing the cost of inserting FA inside the bilayer and ii) a volume term accounting for the dilution of the polymers inside the vesicle due to the increase in area. Because experimental data show that vesicle growth is very small at low internal polymer concentration, we neglect the membrane term and only consider the osmotically-driven contribution to the growth process. Using the Flory-Huggins energy to model the solvent-polymer mixture inside the vesicle, the time evolution of the vesicle area is described by the following nonlinear differential equation (see Supplementary materials):



\begin{equation}
\frac{\partial A}{\partial t} =  \underbrace{\frac{\mathcal{N} a^2 c D A_0^3 \phi^2}{8 \sqrt{\pi }  \ell v_s }}_{\alpha} \frac{1}{A^{3/2}}
\label{eqfinal}
\end{equation}

\noindent where $\phi$ is the initial volume fraction of polymer and $v_s$ the volume of a solvent molecule. This nonlinear differential equation, supplemented by the initial condition $A(0) = A_0$, has the following solution: 

\begin{equation}
\displaystyle{A = \left(A_0^{\frac{5}{2}} +\frac{5}{2} \alpha t \right)^{\frac{2}{5}}}
\end{equation}

In order to check the functional time-dependance of the vesicle area, we plotted the quantity $A^{\frac{5}{2}}-A_0^{\frac{5}{2}}$ for all data curves. As can be seen on Fig.\ref{fig:growth-curves} G, these curves show a clear linear time dependance. The slopes of these curves were then extracted and plotted against $c A_0^3 \phi^2$ (inset of Fig. \ref{fig:growth-curves} G). The linear slope of this curve reveals a nice agreement of the data with the theoretical model and gives  $\mathcal{N} a^2 D / (8 \sqrt{\pi } \ell v_s) = 2.7 \pm 0.2 \times 10^{10} m^5/s$. Using typical values ($a=30$ {\AA}$^2$, D$=10^{-10} m^2/s$, $v_s = 3.10^{-30} m^3$), this yields a value for the thickness $\ell$ of $\sim 2 \mu$m. This result indicates that the relevant lengthscale for FA transport is not the bilayer thickness, as one may think naively, but rather the thickness of the diffusive boundary layer, i.e the lengthscale of the concentration gradient between the bath and the membrane.  At higher FA (or lower PEG) concentration, the hypothesis that fatty acids insertion is the limiting factor of vesicle growth breaks down. In that case, the permeation of water molecules inside the vesicles becomes the limiting factor of the growth process.  While a full model of coupled water and FA transports is outside the scope of this paper, we can estimate the relative influence of these two terms. At high FA concentration, the driving force for FA insertion is the difference in chemical potential between the exterior and the vesicle membrane. Because the latter does not depend on the external FA concentration $c$, this driving force will behave as $c\log{\frac{c}{c_0}}$ at high enough $c$. The osmotic driving force for water molecules on the other hand behaves as $\phi^2$ and we can thus expect that stable growth will occur when $(c\log{\frac{c}{c_0}})/\phi^2$ is below a critical value while growth will become unstable when this ratio is above the critical value. Using $c_0 = 40\mu$M, we find that the curve $(c\log{\frac{c}{c_0}})/\phi^2=14 \mu$M indeed partitions the phase diagram into stable and unstable regions.


%
%
%



\begin{figure}[t!]
	\centering
		\includegraphics[width=0.9\columnwidth]{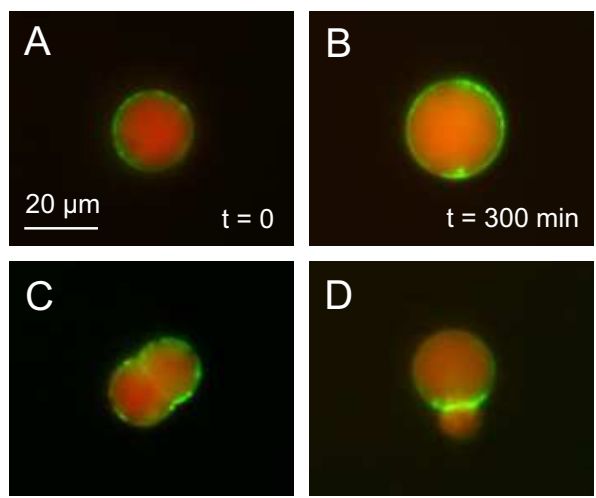}
		\caption{(color online). Growth of vesicles encapsulating a CFER system and expressing the proteins MreB, yfp-MreB and MreC. A) Before growth, the MreB protein (false-colored green) is localized at the membrane surface and the interior of the GUV is tagged with PEG-rhodamine (false-colored red). B) Addition of a mixture of FA led to stable growth of the artificial cell. In the experiment shown here, the vesicle volume is $\sim 2.2$ times its initial volume after 300 minutes of growth. C: and D:  Two examples of budded vesicles after 300 minutes of growth. These initially spherical vesicles have lost their sphericity because of the geometrical constraint due to the network of cytoskeletal protein beneath the membrane.}
	\label{fig:mreb}
\end{figure}

Next, we investigated the feasibility of achieving stable growth on vesicles expressing a set of cytobacterial proteins \cite{maeda12}. To this end, we encapsulated an efficient cell-free expression-reaction (CFER)  system containing all the necessary component for transcription and translation inside the GUV \cite{shin10,chalmeau11}. Two key proteins of the \textit{E. coli} cytoskeleton, MreB and MreC were cloned inside individual plasmids and expressed concurrently. A third plasmid containing a yfp-tagged MreB protein was also expressed to visualize the formation of this simplified cytoskeleton on the vesicle membrane. The vesicle interior was detected by incorporating $100\mu$M of PEG-rhodamine to the CFER. Because oleic acids precipitate at concentration of cations smaller than those required for gene expression, we used a mixture of oleic acid and monoolein (at ratio 2:1) to stabilize the FA micelles. Furthermore, FA insertion inside the GUV critically relies on a pH of $\sim 8.5$. Because gene expression at this pH is sub-optimal, 1 $\mu$M of the pore-forming protein $\alpha$-hemolysin from \textit{Staphylococcus aureus} was added to the external medium to compensate for the sub-optimal pH and enhance protein production inside the GUV. After 12 hours of gene expression, a layer of MreB proteins could be seen on the vesicle membrane (Fig.\ref{fig:mreb}-A). GUVs were then transferred in a solution containing a mixture of FA at $60\mu$M and the volume of these vesicles doubled within 300 minutes as seen on Fig. \ref{fig:mreb}  B. In some instances, the cortex of cytoskeletal proteins prevented the radially symmetric growth of the vesicles and led to the formation of buds  during the growth process (Fig.\ref{fig:mreb} C and D). While several vesicles grew successfully, other vesicles also exploded upon FA addition as did all the vesicles at higher ($\geq 160\mu$M) FA concentrations. Consequently, no unstable growth could be triggered, even when the osmotic pressure difference was lowered by adding PEG to the external medium.

We have

In order to trigger artificial cell division by taking advantage of the unstable growth mechanism described previously, as observed in L-shape bacteria \cite{mercier13}, it will therefore be necessary to find more stable combinations of fatty acids. Alternatively,  an additional forcing can be provided to separate buds from mother vesicles. For example, experiments have shown that contractile forces are produced spontaneously when the FtsZ protein ring is reconstructed inside liposomes \cite{osawa08}. Expressing this protein together with the primitive cytoskeleton described here could provide a possible mechanism of artificial cell division. 



\paragraph{\textbf{Acknowledgments}}
A.L. And V.N. aknowledge support from HFSP account RGP 0037/2015.

\end{document}